\documentclass[aps,pre,amsmath,amssymb,twocolumn,groupedaddress,showpacs]{revtex4-1}
\usepackage{graphicx}
\usepackage[utf8]{inputenc}

\usepackage{epsf}

\newcommand{\be}{\begin{equation}}
\newcommand{\ee}{\end{equation}}
\newcommand{\ba}{\begin{eqnarray}}
\newcommand{\ea}{\end{eqnarray}}
\newcommand{\baa}{\begin{eqnarray}}
\newcommand{\eaa}{\end{eqnarray}}
\newcommand{\ed}{\end{document}}

\newcommand{\re}[1]{(\ref{#1})}
\newcommand{\ci}[1]{\cite{#1}}
\renewcommand{\baselinestretch}{1.2}
\date{\today}
\begin{document}%\large
\title{Classical dynamics and particle transport in kicked billiards}
\author{D.U.Matrasulov,  U.R.Salomov, G.M.Milibaeva, N.E.Iskandarov
\\Heat Physics Department of the Uzbek Academy of Sciences\\
28 Katartal St.,700135 Tashkent, Uzbekistan
}

\begin{abstract}
We study nonlinear dynamics of the kicked particle whose motion is
confined by square billiard. The kick source is considered as
localized at the center of square with central symmetric spatial
distribution. It is found that ensemble averaged energy of the
particle diffusively grows as a function of time. This growth is much
more extensive than that of kicked rotor energy. It is shown that
momentum transfer distribution in kicked billiard is considerably
different than that for kicked free particle. Time-dependence of the
ensemble averaged energy for different localizations of the kick
source is also explored. It is found that changing of localization
doesn't lead to crucial changes in the time-dependence of the
energy.
{\it Also, escape and transport of particles are studied by considering kicked open billiard with one and three holes, respectively.
It is found that for the open billiard with one hole the number of (non-interacting) billiard particles decreases according to exponential law.}
\end{abstract}
%\pacs{05.45.Ac, 05.45.Pq,05.45.Gg.}
\maketitle

\section{Introduction}
~~~Periodically driven dynamical systems play one of the central role in
classical and quantum chaos theory \ci{chir}-\ci{buch02}.
An important feature of  periodically driven dynamical systems is the
chaotization of the motion under certain conditions (resonances, exceeding
critical value of the external field strength etc.).
This chaotization leads to the exponential divergence of neighboring
trajectories in the phase space and diffusive growth of the energy of the given system \ci{chir}-\ci{cas79}.

Comprehensive theoretical \ci{chir}-\ci{buch02} and experimental \ci{moor95}
study of the simplest periodically driven systems kicked rotor shows that for higher enough values of the kicking
force the average energy of the system grows linearly in time. Depending on the kicking strength dynamics of the
system can be mixed or chaotic \ci{chir}-\ci{buch02}.

In this paper we study particle motion in closed and open square billiards in the presence of external
periodic perturbation with the aim to explore the effect of confinement on periodically driven dynamics.
The motion of a particle in confined geometries is a paradigm for the study of
nonlinear dynamics and chaos in theoretical \ci{buch02}-\ci{gutk86} and
experimental contexts \ci{stei92}-\ci{demb01} in case of time-dependent
systems. Up to now much progress is done in the study of billiards with static
boundaries. {\it Also, classical dynamics of the billiards with time-dependent boundaries have been extensively studies in the context of
Fermi acceleration \ci{losk}-\ci{koil}. In particular, possibility of Fermi acceleration in Sinai \ci{losk}, stadium \ci{Losk02} and
elliptical \ci{flor07,flor08} billiards has been shown.}

Dynamics of unperturbed billiard is governed by the geometry of its boundaries, while kicked billiard has additional
factors which allows to manipulate by the dynamics. Those are perturbation parameters
such as coupling constant and frequency of the kick.

We explore the kicked billiard particle dynamics by calculating time dependence of the energy both for
a single trajectory and ensemble of the trajectories. In addition, we treat momentum transfer distribution for this system
and compare it with that of kicked rotor.

Our study shows that depending on the type of the kick potential,
localization and strength of the perturbation force the dynamics can
be different. In particular, the motion of the particle can be localized (trapped) in the kicking area.
The 'life time' of such trap depends on the perturbation strength
(coupling constant) and initial energy of the particle.

It is found that the average energy of the kicked particle grows diffusively as a function of time,
as does the kicked rotor average energy. However, this growth is more rapid than that of kicked rotor.

Also, we study  particle transport in kicked billiard by considering open billiards with one and three holes.
We explore time-dependence of the escape rate and transmission and reflection coefficients.
We note that classical dynamics of unperturbed open billiards have been studied for integrable and chaotic geometries by many authors
(see, e.g., Refs.~\ci{bau90}-\ci{sti07}). In particular, it was found that the number of (non-interacting) particles in
non-integrable open billiard decreases exponentially, while
in case of regular billiard it decreases according to power law \ci{bau90}-\ci{vic01}. In this paper we extent these studies for the case 
when non-interacting particles in an open billiard subjected into the influence of delta-kicking force.

The motivation for the study of periodically driven billiards is caused by several reasons. In many systems with confinement
(e.g. quantum dots, graphene, MIT bag model) in real situations a confined system is subjected to the action of external time-dependent fields.
For example, hadrons in quark-gluon plasma can be considered as driven confined system. In quantum dots or in graphene, external time-periodic
perturbation can be used for manipulating by particle transport in these systems.

This paper is organized as follows. In the next section we will give formulation of the problem for the kicked square
billiard and its detailed solution. Section III extends the results of section II for the case of
kicked open billiard with one and three holes. Last section provides some concluding remarks.

\section{Kicked billiard vs Kicked rotor}
\begin{figure}[htb]
\centerline{\includegraphics[width=6.5cm]{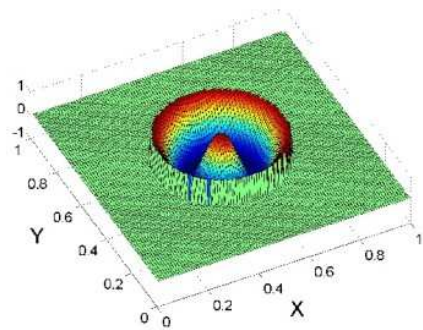}}
\caption{(Color online) Three-dimensional plot of the kicking potential for $\alpha =1.0$.}
\label{fig1}
\end{figure}

In this work, we examine a system consisting of a particle moving inside a two-dimensional square billiard
with an additional kicking source located at the center of the billiard. The kicking potential (see Fig. \ref{fig1}) is given by
\begin{eqnarray} \label{eq1}
V(x,y,t)= \begin{cases} \alpha \cos(\frac{8\pi \rho}{a})\sum_n\delta(t-nT) \quad &\rho\le R \\
0  & \rho>R \\ \end{cases}
\end{eqnarray}
with
$$
\rho=\sqrt{\left(x-\frac{a}{2}\right)^2+\left(y-\frac{a}{2}\right)^2},
$$
where $\alpha, T$, $a$ and $R$ are the coupling constant, the kicking period, the side length of the square and
the radius of kicking area, respectively. Furthermore we fix $R=a/4$.
The Hamiltonian of the system can be written as
\be
H=H_0+V(\rho,t),
\label{eq2}
\ee
with $H_0$ being the Hamiltonian of the particle moving in the square billiard without any kicking. Since particles move balistically
in between collision with the billiard boundary as well as in between kicks, a discrete mapping is used to solve the corresponding equations of motion.
At collisions with the billiard boundary the particles undergo elastic reflections, whereas if the particles are at $t=nT, \, n=1,2,3, \dots$ inside the
kicking area, their momenta changes according to
\begin{equation}
 \vec{p'}= \vec{p} - \vec{\nabla}  V(x,y,t),
\end{equation}
where $\vec{p}$ and $\vec{p'}$ are the momenta just before and right after the kick respectively.
\begin{figure}[htb]
\centerline{\includegraphics[width=6.5cm]{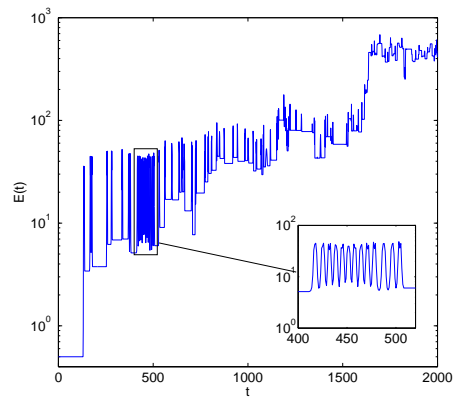}}
\caption{(Color online) Time-dependence of the energy for a typical particle in the kicked billiard ($\alpha =0.2$, $T=0.01$). Inset: rapid oscillations
of the energy, the particle is trapped for a certain time inside the kicking area.}
\label{fig2}
\end{figure}
\begin{figure}[htb]
\centerline{\includegraphics[width=6.5cm]{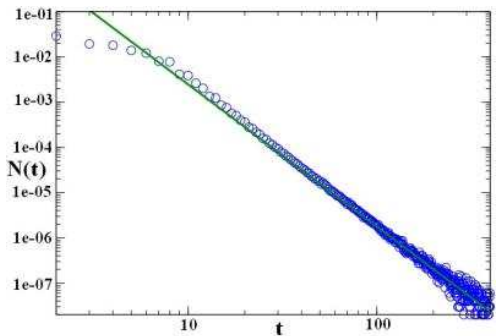}}
\caption{(Color online) Distribution of constant energy time intervals. Circles are the results of direct numerical calculations,
solid line is the linear regression}
\label{fig3}
\end{figure}
In Fig.~\ref{fig2}, the energy $E(t)$ as a function of time for a typical trajectory is shown. The curve can be decomposed in characteristic segments,
each corresponding to a representative dynamics of the particle: Firstly, parts with rapid oscillations of the energy can be seen, see the inset of Fig.~\ref{fig2}.
During such phases, the particle is trapped for certain time inside the kicking area and experiences successive kicks, leading to the typical oscillations of
the energy shown in the inset of Fig.~\ref{fig2}. Secondly, single vertical lines in the curve of Fig.~\ref{fig2} correspond to single kicks, the energy of
the particle is either increased or decreased, depending on where the kick happens. Finally, parts of the curve where the energy remains constant. During such
times, the particle moves outside the kicking area (quasiperiodic orbits) or crosses the kicking area without getting a kick, the latter means it is in a way not synchronized with the kicking period.

To understand the kicked billiard dynamics more deeply we need to explore distribution, $N(t)$ of the time intervals during
which the energy remains constant, i.e. constant-energy-time intervals.
Fig. 3 presents the plot of such distribution in double logarithmic scale obtained using direct numerical computation.
In addition, this figure compares also $N(t)$ with the curve
$N(t)$ obtained from the linear regression.
Such power law (with the exponent equal to $-3$) behavior of $N(t)$ can be explained as follows: appearing constant energy intervals is caused by two factors.
One of them corresponds to the situation when
billiard particle moves along the quasiperiodic  orbits which initially doesn't cross the kicking area. For this case one can obtain the estimate
$N(t)\sim t^{-1}$. However, we found that the probability for appearing such orbits in our system is quite small.
Therefore such regime of motion doesn't make contribution to the above distribution of constant-energy-time intervals. Second type of motion corresponds
to the above mentioned synchronized motion. It is easy to show that the distribution for the constant energy time intervals for this case behaves as $N(t)\sim t^{-3}$.
\begin{figure}[htb]
\centerline{\includegraphics[width=6.5cm]{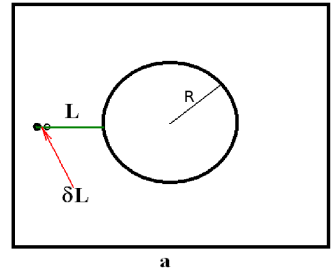}}
\caption{Schematic picture}
\label{sch}
\end{figure}

To explain this we note that in the synchronized regime billiard particle returns to its initial position during the kicking period. If in each initial
condition we will shift its initial position to a infinitesimally small distance $\delta L$ (see Fig.~\ref{sch}) its final (after one kicking period) position
shifts to the same distance $\delta L$ becoming closer to the kicking area.
Then time $t$ after which particle reaches the kicking area can be found from the following relation:
\begin{equation}
\frac{L}{\delta L}=\frac{t}{T}
\label{eq:tT}
\end{equation}
where $L$ is the distance between the particle's initial position and the kicking area.

Thus the number of initial conditions is related to the quantity $\delta L$ as
\begin{equation}
N(\delta L)=\int^{\delta L}_{0}\rho(q)d(q)
\label{eq:ndl}
\end{equation}
where $\rho$ is the density of particles.

Since particles are uniformly distributed, $\rho$ depends on $q$ as
\begin{equation}
\rho(q)=\gamma q^2
\end{equation}
where $\gamma$ is coefficient of proportionality.

Therefore it follows from the eq. \eqref{eq:ndl} that
\begin{equation}
N(\delta L)=\gamma_1 (\delta L)^3
\end{equation}
or, combining with the eq.\eqref{eq:tT} we get finally:
\begin{equation}
N(t)\sim N(\delta L)=\gamma_2 t^{-3}
\end{equation}

\begin{figure}[htb]
\centerline{\includegraphics[width=6.5cm]{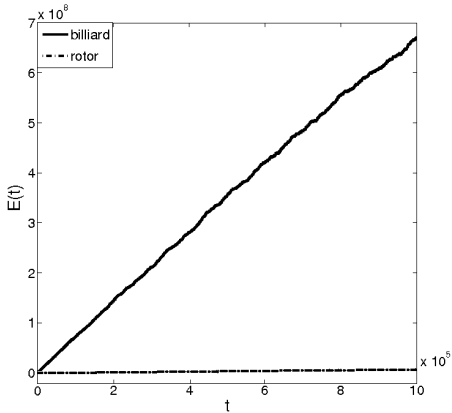}}
\caption{Comparison of the time-dependence of the ensemble averaged energy of the kicked billiard and kicked rotor for $\alpha =5.0$, $T=1.0$.}
\label{fig4}
\end{figure}

When considering not only a single trajectory, but rather an ensemble of particles, a diffusive growth of the ensemble averaged energy
$\langle E_b(t) \rangle$ (averaged of 1000 trajectories) can be observed, see Fig.~\ref{fig4}. More precisely, the energy growth linearly 
(normal diffusion) with the time $t$. The proportionality constant is naturally just the diffusion coefficient $D_b(\alpha, T)$, so that
$\langle E_b(t) \rangle = D_b(\alpha, T) \cdot t$. As indicated, $D_b$ depends on the coupling constant $\alpha$ and the kicking period $T$.
$D_b$ increases monotonically with increasing $\alpha$ and decreases monotonically with increasing $T$, see Fig.~\ref{fig6}.

\begin{figure}[htb]
\centerline{\includegraphics[width=6.5cm]{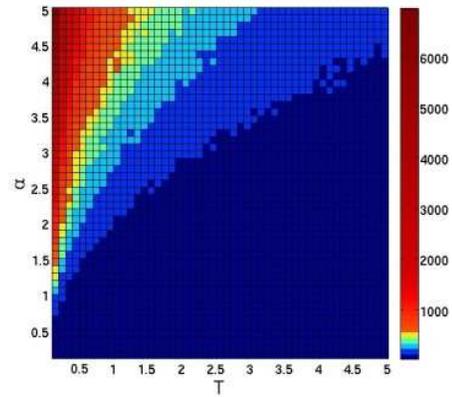}}
\caption{(Color online) The dependence of the diffusion coefficient $D_b$ on  $\alpha$ and $T$ for  the kicked
 billiard.}
\label{fig5}
\end{figure}

\begin{figure}[htb]
\centerline{\includegraphics[width=6.5cm]{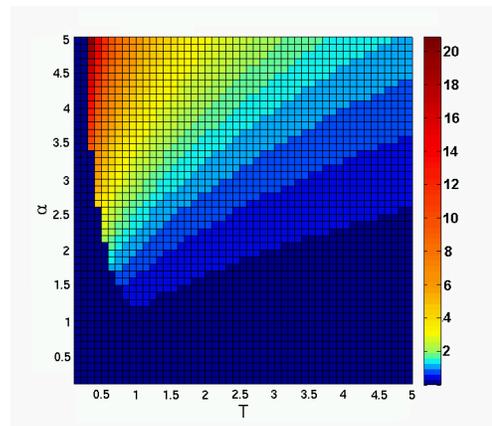}}
\caption{(Color online) The dependence of the diffusion coefficient $D_r$ on  $\alpha$ and $T$ for  the kicked
 rotor.}
\label{fig6}
\end{figure}

It is reasonable to compare the diffusive growth of the ensemble averaged energy $\langle E_b(t) \rangle = D_b(\alpha, T) \cdot t$ of the 
kicked billiard with the evolution of energy $\langle E_r(t) \rangle$ of the kicked rotor (again with $\alpha$ being the coupling constant 
and $T$ being the period between two successive kicks). From Fig~\ref{fig4}, it can be seen that $\langle E_r(t) \rangle =D_r(\alpha, T) \cdot t$,
but with $D_r(\alpha, T) \ll D_b(\alpha, T)$, so the energy growths much faster in the case of the kicked billiard. We note that in the case
of the kicked rotor the dynamics is effectively governed by a single parameter $K_r=\alpha T$ only, see e.g \cite{Izr90}. The energy of the
kicked rotor as a function of the dimensionless time $n=t/T$ can than for $K_r\gtrsim5$ be written as $E_r(n)= K_r^2/4 \cdot n$, so
\begin{equation}
\label{eq:Dr}
D_r(\alpha, T) = D_r(K_r) = K_r^2/4.
\end{equation}
In the kicked billiard the situation is different, the dynamics depends on $\alpha$ and $T$ individually, thus there is no simple representation
for $D_b(\alpha, T)$ as in Eq. \eqref{eq:Dr} possible. The dependence of $D_b(\alpha, T)$ on $\alpha$ and $T$ is shown in Fig. ~\ref{fig4},
whereas $D_r(\alpha, T)$ is shown in Fig.~\ref{fig5}. The large deviation between $D_b$ and $D_r$ becomes immediately clear when considering
the maximum momentum transfer $\Delta P_{max}$ at a single kick. In the case of the kicked billiard, $\Delta P_{max}=8\pi\alpha/a$, whereas
for the kicked rotor, $\Delta P_{max}= \alpha$.

\begin{figure}[htb]
\centerline{\includegraphics[width=6.5cm]{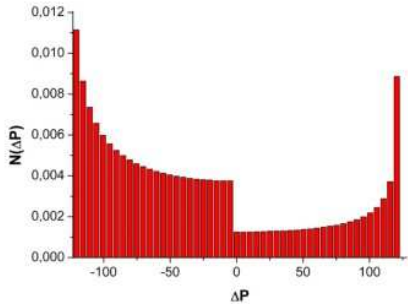}}
\caption{(Color online) Distributions of momentum transfers $\Delta P$ for the kicked billiard for $\alpha=5.0$ and for $T=1.0$;}
\label{fig7}
\end{figure}

\begin{figure}[htb]
\centerline{\includegraphics[width=6.5cm]{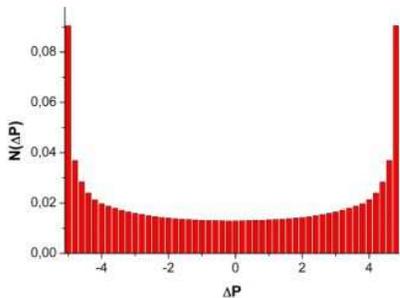}}
\caption{(Color online) Distributions of momentum transfers $\Delta P$ for the kicked rotor for $\alpha =5.0$ , $T=1.0$.}
\label{fig8}
\end{figure}

The corresponding distributions $N_r(\Delta P)$ and $N_b(\Delta P)$ (1000 initial conditions iterated until $t=10^6 T$) are shown in
Figs.~\ref{fig7},~\ref{fig8} (the subscripts $r$ and $b$ denote the kicked rotor, the kicked billiard respectively). Clearly, the available
range in $\Delta P$ increases with increase of $\alpha$ in the case of the kicked billiard, and for a fixed $\alpha$ this range is much larger
than in the case of the kicked rotor. Unlike the $N_r(\Delta P)$, the distribution $N_b(\Delta P)$ has a jump at $\Delta P=0$, this can be 
explained as following.
The momentum transfer in the case of kicked rotor can be found from 2-D standard map as $\Delta P_r=K_r\sin(\theta)$.
For the kicked billiard the momentum transfer can be written as:
\begin{equation}
\Delta P=|\vec {P'}-\vec P|=K_b\sin(b\rho)
\label{vb}
\end{equation}
where $\vec P$ and $\vec{P'}$, which are the momenta before and after the kick, respectively and $b=8\pi/a$, $K_b=b\alpha$.

Hence, for the momentum transfer distribution we get
\begin{equation}
N_b(\Delta P)\sim N(\rho) \left(\arcsin(\frac{\Delta P+\Delta \delta P}{K})-\arcsin(\frac{\Delta P}{K})\right)
\end{equation}
In the case of kicked rotor we have $N_r(\theta)=const$, for $K_r\gtrsim5$ (chaotic regime), while for kicked billiard
as it was mentioned above, we have equally distributed kick occurrence over $x,y$. Therefore, the dependence of this distribution
on $\rho$ is given by $N_(\rho)=const_1+const_2\rho$. It follows from the eq.(\ref{vb}) that $\Delta P=0$ for $\rho=0$
and for $\rho=a/4$. However, the kick numbers at these two points are not equal, number of kicks at $\rho=a/4$ points
is maximal, while for $\rho=0$ it is minimal that explains the difference between the shape of $N(\Delta P)$ in Figs.
\ref{fig7} and \ref{fig8}.\\
\begin{figure}[htb]
\centerline{\includegraphics[width=6.5cm]{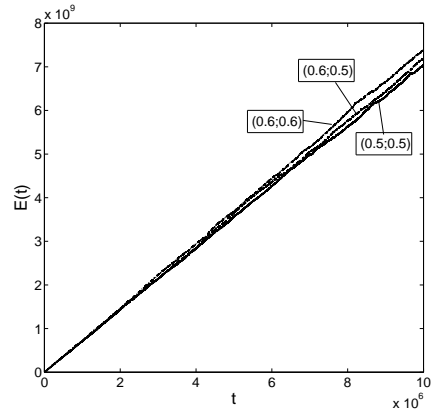}}
\caption{Time-dependence of the ensemble averaged energy for different localizations of
 the center of the kicking source in the billiard  ($\alpha =5$, $T=1$):
The central localization is compared to the case, when kicking
 source is shifted to the left (position of the center at $(0.6; 0.5)$) and when it is shifted along the
diagonal of billiard (center at $(0.6; 0.6)$).}
\label{fig9}
\end{figure}

In Fig. \ref{fig9}, the time-dependence of the average energy $\langle E(t) \rangle$ is plotted for  different localizations of the kicking
source and compared to the one when kicking source is located at the center of the billiard. The shift of the kick source leads to minor changes
in  $\langle E(t) \rangle$ only, so the dynamics is rather robust against this shift. We tested this for other delocalizations as well, with similar
results, thus there are not shown here.

So far, we have considered the kicked billiard with a potential which is very similar to the one of the kicked rotor, in particular with the same
minimum and maximum values of the kicking potential. Despite this, however, the acceleration in the kicked billiard is much more pronounced compared
to the kicked rotor, which is of course due to the fact that much larger $\Delta P$ are possible upon single kicks in the billiard. To make the two
systems better comparable in terms of the momentum transfer, we modify the potential given in Eq. \eqref{eq1} slightly and obtain:
\begin{figure}[htb]
\centerline{\includegraphics[width=6.5cm]{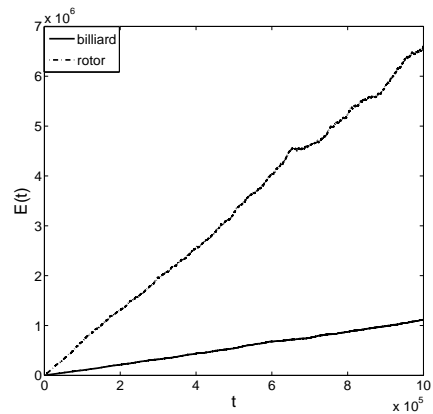}}
\caption{Comparison of the time-dependence of the ensemble averaged energy of the kicked billiard and the kicked rotor with $\alpha =5.0$, $T=1.0$.}
\label{fig10}
\end{figure}
\begin{eqnarray} \label{eq:potential2}
V(x,y,t)= \begin{cases} \frac{\alpha a}{8\pi}\alpha \cos(\frac{8\pi \rho}{a})\sum_n\delta(t-nT) \quad &\rho\le R \\
0  & \rho>R \\ \end{cases}
\end{eqnarray}
This potential leads to the same maximum and minimum momentum transfer as in the kicked rotor. Now, the energy of the kicked rotor growth faster
than the one the kicked billiard, see Figs.~\ref{fig10}. These reasons are the following:
\begin{figure}[htb]
\centerline{\includegraphics[width=6.5cm]{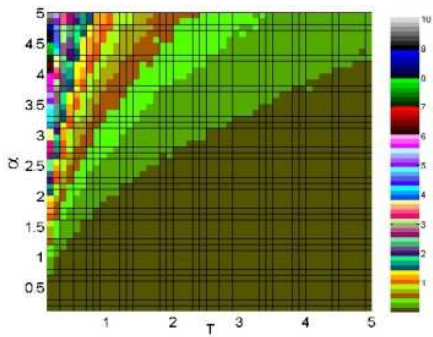}}
\caption{(Color online) The dependence of the diffusion coefficient $D_b$ on $\alpha$ and $T$ for the  kicked
 billiard.}
\label{fig11}
\end{figure}

Firstly, in the kicked rotor at every $t= nT, \, n=1,2,3,\dots$ the particle experiences a kick, a certain amount of momentum is transferred.
In the kicked billiard this is not the case, a particle gets a kick only if is at $t= nT, \, n=1,2,3,\dots$ inside the kicking area, see Fig.~\ref{fig1}.
Consequently, particles will experience fewer kicks during a certain time interval compared to the kicked rotor.

{\it We note that the above considered kicked billiard to some extent is equivalent to time-dependent
Sinai scattering billiard. However, unlike usual Sinai billiard the scatterer is not static and switches on
periodically. In addition, in kicked billiard collision with the scatterer is inelastic, while in Sinai billiard
collision with the scatterer is elastic. Here we mention that time-dependent Sinai billiard has been studied earlier
in the context of Fermi acceleration \ci{losk,Losk02,Peter1}. However, these studies deal with the
billiards whose boundaries are time-dependent, while in our system the boundaries of the billiard are static.

Therefore exploring of such time-dependent billiards requires solving the classical equations
of motion with time-dependent boundary conditions.
Time-dependence in our system is caused by the kicking source and we do not need to consider time-dependent boundary conditions
in the equations of motion.

Also, as it was shown, under certain conditions particle motion in the kicked billiard can be
localized (trapping of particle into the kicking source) in the kicking area. Such a trapping is not possible for usual
time-dependent Sinai billiard.

Finally, it should be noted that particle acceleration in kicked billiard can be considered as a kind of Fermi acceleration.
In usual Fermi acceleration the growth of the kinetic energy of a billiard particle is caused by the motion of billiard walls,
while in our case the energy grows due to the interaction with the kicking source.
The latter mechanism is close to the kicked rotor rather than to the billiard with time-dependent boundaries.
Detailed study of Fermi acceleration in billiard geometries showed that the acceleration is possible
for non-integrable billiards only \ci{losk,Losk02}. However, the above studied kicked square billiard has integrable boundaries. Despite this
acceleration is possible in this system. This is the main difference between the above kicked billiard and other time-dependent billiards 
(with time-changing boundaries).}

\section{Open billiard}
\begin{figure}[htb]
\centerline{\includegraphics[width=6.5cm]{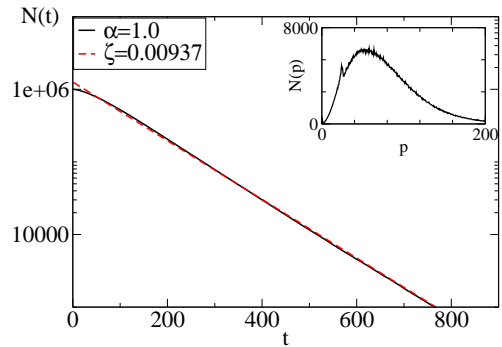}}
\caption{(Color online) Time-dependence of the number of survived particles $N(t)$ for the value $\alpha=1.0$
and calculated using eq.~\re{sur:eq}. $\sigma=0.001 \times a, N_0=10^6, p_0=1.0$.
The inset: Distribution of the escaped particle momenta.}
\label{a1c}
\end{figure}
\begin{figure}[htb]
\centerline{\includegraphics[width=6.5cm]{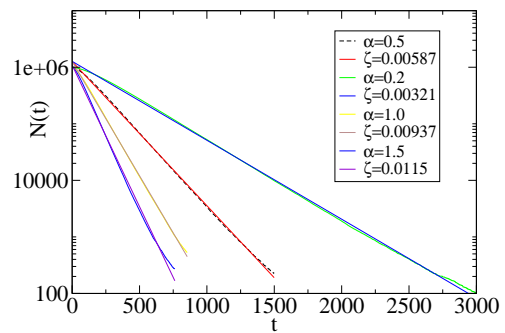}}
\caption{(Color online) Time-dependence of the number of survived particles $N(t)$ for different values of the delta-kick strength
$\alpha$.$\sigma=0.001 \times a, N_0=10^6, p_0=1.0$.}
\label{multi}
\end{figure}
In this section we study particle transport in a kicked open billiard.
The first system we will consider is a square billiard having one small hole with the size $\sigma$.
The type and location of the kicking source is the same as in the previous section.
Simulation of the system is performed for $N_0$ {\it non-interacting} particles with randomly distributed initial positions, $(x_0, y_0)$, 
and directions of the initial velocities, $\phi$. The initial momenta of all particles are assumed to be equal $p_0$.
Classical dynamics of unperturbed open billiards has been extensively
studied both for integrable and chaotic geometries \ci{bau90}-\ci{sti07}.
In particular, it was found in the Refs.~\ci{bau90}-\ci{vic01} that for the case of ergodic particle motion in
an open billiard the particle's escape rate decreases exponentially, while for a non-chaotic
system it decreases according to power law.

In this work we explore behavior of the escape rate for the kicked open billiard.
The quantity we want to calculate is the number of particles in billiard at the time t, $N(t)$.
The escape rate is related to $N(t)$ as
$$
\rho(t) = 1-N(t)/N_0 ,
$$
where $N_0$ is the number of particles at $t = 0$.
As it was shown recently \cite{bau90,alt96}, for chaotic billiards time-dependence of survival probability
can be written as
\begin{equation}
\frac{N(t)}{N_0}= \exp(-\zeta t),
\label{sur:eq}
\end{equation}
where
$$
\zeta=\frac{p \sigma}{\pi S}.
$$
with $p$ being the absolute value of the billiard particle's momentum, $S$ is the area of the billiard.
Intuitively, one may expect similar decay in the case of kicked billiard (which is chaotic for arbitrary geometry of billiard boundaries).
However, in this case momenta of billiard particles are not equal and $p$ in eq. ~\re{sur:eq}
can be replaced by $p'$, most probable value of the
escape momentum. In Fig.~\ref{a1c} we compare $N(t)$ computed: \\
i) using eq.~\re{sur:eq} and \\
ii) by numerical modeling for three values of the kicking strength, $\alpha =1.0$. \\
As it can be seen from these plots, increasing of the kicking force leads
to decreasing of $N(t)$ which is equivalent to increasing of the number of escaped particles.
Such behavior can be explained
by the fact that according to the previous section, for higher values of the kicking force the energy of the billiard particles
growth linearly in time.
Also, the plots in Fig.~\ref{multi} show that the results of simulation are in good agreement with the results obtained using
eq.~\re{sur:eq}.

\begin{figure}[htb]
\centerline{\includegraphics[width=6.5cm]{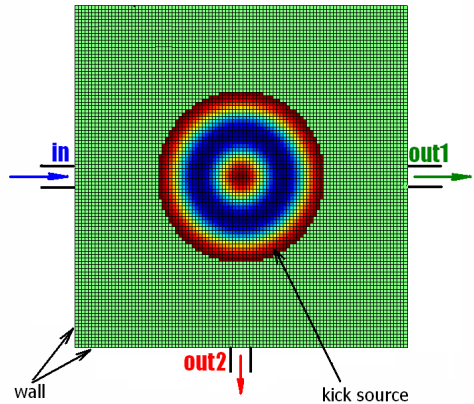}}
\caption{(Color online) Kicked square billiard with three holes}
\label{leads}
\end{figure}

\begin{figure}[htb]
\centerline{\includegraphics[width=6.5cm]{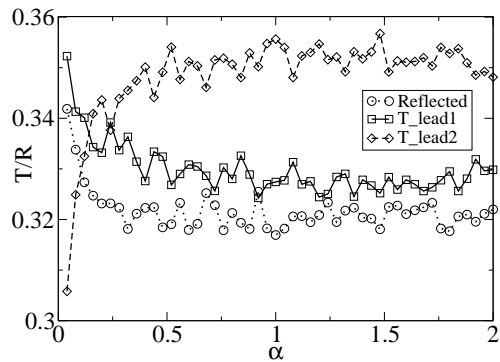}}
\caption{Transmission ($T$) and reflection ($R$) coefficients versus kicking force $\alpha$.
$\sigma=0.001 \times a, N_0=10^6, p_0=1.0$}
\label{fig13}
\end{figure}

To study transport properties of the kicked billiard system we consider square billiard having three holes
with attached one incoming and two outgoing leads Fig.~\ref{leads}. Particles are assumed to come into
the billiard from incoming lead, while their escape from the billiard is possible from all three leads. Reflection coefficient
$R$ is calculated as the escape rate from the incoming lead, while transmission coefficients $T_1$ and $T_2$ are the escape rates from the outgoing leads.
In Fig.~\ref{fig13} these coefficients are plotted as a function of kicking strength. As it can be seen from these plots, the
reflection coefficient is much smaller compared to transmission coefficients for the leads 1 and 2.
This implies that such system can be used as a conductance amplifier. All the
coefficients become $\alpha-$independent for the values of $\alpha$ larger than $0.5$.

\section{Summary}
In this work we studied classical dynamics of a kicked particle whose motion is confined in a square billiard.
The kick potential is considered as localized inside the billiard with central symmetric spatial distribution.
It is found that for this type of kick potential the average energy of the particle as a function of time grows faster than that of kicked free particle.
This implies that the above explored system is more attractive from the view point of acceleration.
Unlike the kicked free particle (kicked rotor) dynamics of the kicked billiard particle depends of perturbation parameters $\alpha$ and $T$ separately.
It also depends on the geometry of billiard and localization of the kick source.
In addition, the case of the kick potential that gives the same maximum and minimum values of momentum transfer as those for kicked rotor is considered.
It is found that in this system the acceleration is much weaker compared to the first system and even compared to kicked rotor.

Also, we have studied particle transport in kicked open billiard with one and three holes. In particular, we calculated escape rate, transmission
and reflection coefficients. The latter quantities are calculated for a billiard with one incoming and two outgoing holes.
Exponential decay {\it (i.e., decreasing the number of billiard particle according to exponential law)} of the system is found
for the case of one-hole open billiard. It is shown for such billiard that increasing
of the kicking force leads to rapid decay of the system. Transmission and reflection probabilities are studied for the case of a kicked billiard with
three wholes. For this system injection of particles into the billiard occurs from one hole, while escape of particles occurs from all three holes.
It is shown that for such system transmission and reflection coefficients depend on the kick strength, $\alpha$ until we increase $\alpha$ 
up to certain value. Further increasing of $\alpha$ does not lead to changes in $T$ and $R$.

{\it Finally, we note that the above considered kicked billiard is a kind of time-dependent billiards whose dynamics are
completely different than that of static billiards. Recently billiards with time-dependent boundaries have been extensively studied
\ci{losk} -\ci{koil} in the context of Fermi acceleration. In such billiards the dynamics is governed by the
time-changing law of the boundaries. This time-dependence requires solving of classical equations of motion with time-dependent boundary conditions.
Unlike these billiards, the kicked billiard is perturbed by the kicking source located inside the billiard
and the walls of the billiard are static. This fact causes the main difference in the dynamics of the kicked billiards and
the billiards with moving walls.}

The importance of the above study is caused by its perspective relevance to the particle transport
in various mesoscopic systems (e.g. as quantum dots, ratchets, nanotubes etc.).
Time-dependent external field can be used in these systems as an additional tool for manipulating by particle dynamics and
transport. In this context an important issue should be the extension of the above study for the case of corresponding
quantum system. Such a study is currently in progress.
\begin{acknowledgments}
This work is supported in part by the grants of the Uzbek Academy of Sciences (FA-F2-084, FA-F2-082)
and the grant of Volkswagen Foundation (Nr I/80 136).
\end{acknowledgments}

%\newpage

\end{document}